\documentclass[]{aa}
\usepackage[varg]{txfonts}
\usepackage{amsmath,amssymb}

\newcommand{\Teff}{\ensuremath{T_\mathrm{eff}}}
\newcommand{\zav}[1]{\left(#1\right)}

\begin{document}

\title{Hot-star wind models with magnetically split line blanketing}

\author{J.~Krti\v{c}ka}

\institute{\'Ustav teoretick\'e fyziky a astrofyziky, Masarykova univerzita,
           Kotl\'a\v rsk\' a 2, CZ-611\,37 Brno, Czech
           Republic}

\date{Received}

\abstract{Fraction of hot stars posses strong magnetic fields that channel their
radiatively driven outflows. We study the influence of line splitting in the
magnetic field (Zeeman effect) on the wind properties. We use our own global
wind code with radiative transfer in the comoving frame to understand the
influence of the Zeeman splitting on the line force. We show that the Zeeman
splitting has a negligible influence on the line force for magnetic fields that
are weaker than about 100~kG. This means that the wind mass-loss rates and
terminal velocities are  not affected by the magnetic line splitting for
magnetic fields as are typically found on the surface of nondegenerate stars.
Neither have we found any strong flux variability that would be due to the
magnetically split line blanketing.}

\keywords{stars: winds, outflows -- stars:   mass-loss  -- stars:  early-type --
stars: magnetic field -- hydrodynamics -- radiative transfer}

\maketitle

\section{Introduction}

The surface magnetic fields of about 10\% of hot spectral type A and late-B stars have strengths on the order of $0.1-10\,$kG \citep{dvojka,rompreh}.
In such stars, the radiative diffusion may operate in a relatively quiet
environment, leading to chemical peculiarity \citep{vaupreh,mpoprad}. Precise
spectropolarimetric observations show that about the same fraction of O and
early-B stars also have strong magnetic fields \citep{morbob,wamimes,grunmimes}.
In these stars the radiative force launches mass outflow, that is, the stellar wind
\citep[see][for a review]{pulvina} that allows for interaction between the magnetic
field and the wind.

The radiatively driven wind of hot stars is ionized, therefore it flows along
the magnetic field lines. That the stellar wind is channeled along the magnetic
field has numerous observational consequences \citep{malykor}. When
the stellar wind energy density dominates the magnetic field energy
density, the magnetic field opens up and the wind leaves the star \citep{udo}.
The opposite case leads to relatively complex flow structures that include the
inhibition of the outflow and fall-back of the wind onto the stellar surface
\citep{udorot,kuk}, or the trapping of the wind in centrifugally supported
clouds \citep{labor,towog}.

The interaction of the stellar wind with a strong magnetic field has evolutionary
consequences. The wind is forced to the corotation at large distances from the
star, leading to angular momentum loss and rotational braking
\citep{brzdud,membrzd}. This effect was discovered not only on evolutionary
timescales \citep{shauc}, but also on human timescales \citep{town}. Moreover, that the stellar wind is channeled by the magnetic field also affects the mass-loss
rate. The local wind mass flux becomes proportional to the tilt of the magnetic
field \citep{owoudan}. Moreover, wind may leave a star only along open magnetic
field lines, but it falls back along closed magnetic field lines \citep{owoan}.
The resulting wind quenching leads to an additional reduction of the mass-loss rate
that resembles a weakening of the wind at low metallicity. This means that magnetic
stars lose less mass than their non-magnetic counterparts, and the magnetic
fields provide an alternative explanation of the high mass of black hole binary
merger progenitors \citep{magvln}.

The magnetic field affects not only wind dynamics, but also the radiative
transfer, which may be important in radiatively driven winds. The Zeeman and
Hanle effects lead to the polarization of the radiation in spectral lines. This
might be used to detect even relatively weak magnetic fields in the winds
\citep{igzeeman,ihanle,gizeeman}. Moreover, the associated line splitting affects the line force and therefore also the mass-loss rate. Stronger
absorption due to line splitting may enhance the wind blanketing effect
\citep{acko,hd191612}, which contributes to the light variability that is observed in
magnetic O stars \citep{koeyer,nazdis}.

Despite its possible evolutionary consequences, the influence of the Zeeman
effect on line-driven winds has never been studied in greater detail. In
general, this would require self-consistent wind models with polarized line
transfer \citep[e.g.,][]{adam} that account for the mutual radiative interaction of
individual Zeeman components induced by the Doppler effect
\citep{igzeeman,gizeeman}. Such models are not available. However, the strongest
influence of the Zeeman effect on the line-driving mechanism presumably arises
from the line splitting, which may modify the line force. Even including
this effect, however, requires wind models for which the radiative force is calculated in a more
advanced approach than with the single-line Sobolev approximation. 

\begin{table*}
\caption{Relative strengths of magnetically split lines. Here 
$\Delta J=J_u-J_l$ and $M_i=-J_u,\dots,J_u$.}
\label{sobel}
\centering
\begin{tabular}{clll}
\hline
$\Delta J$ & $S_i(0)$ for $\Delta M=0$& $S_i(1)$ for $\Delta M=1$ &
 $S_i(-1)$ for $\Delta M=-1$\\
\hline
0 & $M_i^2$ & $\frac{1}{4}(J_u+M_i)(J_u+1-M_i)$
  & $\frac{1}{4}(J_u-M_i)(J_u+1+M_i)$ \\[2pt]
1 & $J_u^2-M_i^2$ & $\frac{1}{4}(J_u+M_i)(J_u-1+M_i)$
  & $\frac{1}{4}(J_u-M_i)(J_u-1-M_i)$ \\[2pt]
$-1$\,\,\,\,& $(J_u+1)^2-M_i^2$ & $\frac{1}{4}(J_u+1-M_i)(J_u-M_i+2)$ 
  & $\frac{1}{4}(J_u+1+M_i)(J_u+M_i+2)$\\[2pt]
\hline
\end{tabular}
\end{table*}

While the dynamical effects of the magnetic field (i.e., the magnetic field tilt and
the field divergence) on the mass-loss rate have been studied in detail using
magnetohydrodynamic (MHD) models \citep{udo}, the effect of the line splitting was neglected. This
might have a significant effect on the reliability of evolutionary models that
include magnetized mass-loss \citep[e.g.,][]{magvln}. To understand the
influence of the Zeeman effect on the radiative force and on the wind mass-loss
rate, we modified our METUJE wind models to account for Zeeman splitting. Our
wind models calculate the radiative force consistently in the comoving frame
(CMF) in a global approach. In this way, the models account for the interaction of
individual Zeeman components and allowed us to predict the influence of
magnetically split line blanketing on emergent fluxes. To pinpoint the effect of
the line splitting, we neglect the dynamical effects of the magnetic field
connected with wind channeling along the magnetic field lines.

\section{Global wind models}

Wind models with magnetically split line blanketing were calculated using the
METUJE code \citep{cmfkont}. The code provides global (unified) models of the
stellar photosphere and radiatively driven wind. The METUJE code solves the
radiative transfer equation, the kinetic (statistical) equilibrium equations,
and the equations of continuity, momentum, and energy in the photosphere and in
the wind. Models are calculated assuming stationary (time-independent) and
spherically symmetric wind flow.

The radiative transfer equation is solved in the comoving-frame
\citep[CMF,][]{mikuh}. To solve the equation, we account for line and continuum
transitions that are relevant in photospheres and winds of hot stars. The considered
elements and ions are listed in \citet{nlteiii}.

The ionization and excitation state is calculated from the kinetic equilibrium
equations (also called non-local thermal equilibrium (NLTE) equations, see \citealt{hubenymihalas}). We account
for the radiative and collisional excitation, deexcitation, ionization, and
recombination. The bound-free radiative rates are consistently calculated from
the CMF mean intensity, while the bound-bound rates rely on the Sobolev
approximation. The ion models were either adopted from the TLUSTY model
stellar atmosphere input data \citep{ostar2003,bstar2006} or prepared by us.
Both sources use the same strategy to construct the ionic models, that is, the
data are based on the Opacity and Iron Project calculations \citep{topt,zel0}
and are corrected for the observational line and level data available in the NIST
database \citep{nist}. An exception is the ionic model of phosphorus, which was
prepared using data described by \citet{pahole}. The ionic levels with low
excitation energy are explicitly included in the calculations, while levels
with higher excitation energy are merged into superlevels \citep[see][for
details]{ostar2003,bstar2006}.

Depending on the location in the atmosphere, we use three different methods to
solve the energy equation. The differential form of the transfer equation is
applied deep in the photosphere, while the integral form of this equation is
used in the upper layers of the photosphere \citep{kubii}, and the electron
thermal balance method \citep{kpp} is applied in the wind. In all three cases,
the individual terms in the energy equation are taken from the CMF radiative
field. These terms, together with the CMF radiative force calculated accounting for
line, bound-free, and free-free transitions and light scattering on free
electrons, are inserted in the hydrodynamical equations. The hydrodynamical
equations, that is, the continuity equation, equation of motion, and the energy
equation, are solved iteratively to obtain the wind density, velocity, and
temperature structure. The final model is derived by varying the base velocity
to search for a smooth transonic solution with the maximum mass-loss rate
\citep{cmfkont}. 

The output from TLUSTY model stellar atmospheres \citep{ostar2003,bstar2006} was
used as the initial guess of the solution in the photosphere. These TLUSTY
models were calculated for the same effective temperature, surface gravity, and
chemical composition as the wind models, but neglecting the magnetic field.

\section{Including magnetically split line blanketing}

The inclusion of the magnetic line splitting into our wind code closely follows
the quantum mechanical theory of the Zeeman effect \citep{sobel,novyzak}. When a magnetic field is present, each atomic level $k$ described by the total, orbital,
and spin angular momentum quantum numbers $J_k$, $L_k$, and $S_k$ is split into
$2J_k+1$ sublevels with magnetic quantum numbers $M_i=-J_k,\dots,J_k$. According
to the selection rules, only the transitions with $\Delta M=M_u-M_l=-1,\,0,\,1$
are allowed between magnetically split upper $u$ and lower $l$ levels. The
splitting of the energy levels leads to the wavelength shift $\Delta\lambda$
relative to the laboratory line wavelength $\lambda_0$
\begin{equation}
\label{zeemstep}
\Delta\lambda=\frac{e\lambda_0^2B}{4\pi m_\text{e}c^2}(g_lM_l-g_uM_u),
\end{equation}
where $e$ and $m_\text{e}$ are the elementary charge and the electron mass, $B$
is the field modulus, and $g_l$ and $g_u$ are the Land\'e factors.

In our non-magnetic models, the line force is calculated based on line
data derived from the VALD database (Piskunov et al. \citeyear{vald1}, Kupka et
al. \citeyear{vald2}) with some updates using the NIST data \citep{nist}. To
account for the magnetic field, we replaced the original lines by their split
components selected according to quantum-mechanical rules and with wavelength
shifts given by Eq.~\ref{zeemstep}. The oscillator strengths of each split line
$j$ were computed from the original oscillator strength~$g\!f$ 
\begin{align}
(g\!f)_j=&\frac{1}{2}S_j(0)(g\!f), \quad\text{for}\;\Delta M=0,\\
(g\!f)_j=&\frac{1}{4}S_j(\pm1)(g\!f), \quad\text{for}\;\Delta M=\pm1,
\end{align}
where the relative line strengths given in Table~\ref{sobel} are additionally
normalized to unity for each group of the Zeeman components
\begin{equation}
\sum_iS_i(-1)=\sum_iS_i(0)=\sum_iS_i(1)=1.
\end{equation}

The Land\'e factors were mostly taken from the Kurucz line
list\footnote{http://kurucz.harvard.edu} using cross-matching of lines with the VALD
line list. For the remaining lines, the Land\'e factors were computed assuming LS
coupling
\begin{equation}
\label{glande}
g_k=1+\frac{J_k(J_k+1)-L_k(L_k+1)+S_k(S_k+1)}{2J_k(J_k+1)}
,\end{equation}
with term designation from the Kurucz line line list, or we assumed mean
Land\'e factors $g_k=1.2$ \citep[e.g.,][]{koks} when the designation was not
available. The number of unsplit lines in the original line list with different
sources of Land\'e factors and the total number of magnetically split lines
that are accounted for in the calculation is given in Table~\ref{magcar}. 

\begin{table}
\caption{Number of unsplit lines in the input line list with different sources of
Land\'e factors (upper rows) and the total number of magnetically split
line components used in calculations (last row).}
\label{magcar}
\centering
\begin{tabular}{lr}
\hline
Number of lines with known Land\'e factors & 140\,474\\
Number of lines with Land\'e factors calculated & 42\,556 \\
\quad assuming LS coupling Eq.~\eqref{glande} \\
Number of lines with assumed $g_k=1.2$ & 35\,468\\\hline
Total number of magnetically split lines & 3\,420\,041\\
\hline
\end{tabular}
\end{table}

\newcommand\ts[1]{\times{10^{-#1}}}
\begin{table*}
\caption{Adopted stellar parameters of the studied stars and derived wind mass-loss
rates}
\label{ohvezpar}
\centering
\begin{tabular}{ccccccccc}
\hline
\hline
$\Teff$ $[\text{K}]$ & $R_{*}$ $[\text{R}_{\odot}]$ & $M$ $[\text{M}_{\odot}]$ &
$\log(L/L_\odot)$ & $Z/Z_\odot$ &
\multicolumn{4}{c}{$\dot M$ $[\text{M}_{\odot}\,\text{yr}^{-1}]$} \\
&&&&& $B=0$\,G & $B=10^3\,$G & $B=10^4\,$G & $B=10^5\,$G\\
\hline
30000& 6.6&12.9&4.50&1.0&$9.26\ts{9}$&$9.26\ts{9}$&$8.72\ts{9}$&$8.01\ts{9}$\\
     &    &    &    &0.5&$4.57\ts{9}$&$5.25\ts{9}$&$5.47\ts{9}$&$3.75\ts{9}$\\
37500&19.8&48.3&5.84&1.0&$1.03\ts{6}$&$1.03\ts{6}$&$1.04\ts{6}$&$0.98\ts{6}$\\
     &    &    &    &0.5&$6.46\ts{7}$&$6.56\ts{7}$&$6.43\ts{7}$&$6.08\ts{7}$\\
42500&18.5&70.3&6.00&1.0&$1.79\ts{6}$&$1.79\ts{6}$&$1.80\ts{6}$&$1.57\ts{6}$\\
     &    &    &    &0.5&$8.54\ts{7}$&$8.59\ts{7}$&$8.53\ts{7}$&$7.53\ts{7}$\\
\hline
\end{tabular}
\end{table*}

The magnetic field varies with radius according to $\text{the div}\boldmath{B}=0$
constraint. We neglected this effect and assumed a constant magnetic field
throughout the whole computational domain, because the mass-loss rate is
determined close to the star, where the magnetic field is nearly equal to its
surface value. Moreover, the magnetic field is typically so strong that it
dominates even in deep photospheric layers, therefore we can assume that the
field has the same strength for great and small optical depths. This enabled us to
split the lines in the external file and not in the code itself, while the
effect of this assumption on our final results is negligible. Moreover, the
magnetic field also varies across the stellar surface. By neglecting these
variations, we provide in fact models for concentric cones, in which the surface
variations of magnetic field can be neglected.

\section{Hot-star wind models with Zeeman line splitting}

The adopted stellar parameters, that is, the effective temperature $\Teff$, radius
$R_{*}$, mass $M$, and luminosity $L$, together with derived mass-loss rates
$\dot M,$ are given in Table~\ref{ohvezpar}. We selected a representative sample
of O-star parameters that correspond to a main-sequence star with
$\Teff=30\,\text{kK}$ and to two supergiants with $\Teff=37.5\,\text{kK}$ and
$\Teff=42.5\,\text{kK}$. The stellar parameters were derived using the formulas of
\citet{okali}. The magnetic field strengths we selected cover typical surface fields
found in O stars, which are up to few kilogauss
\citep{donthetoric,donhd191612,wadhd14,wadngc,hubcpd}. We assumed two metallicity values that correspond to that of our Sun \citep{asp09} and to that of the Large Magellanic
Cloud ($Z=0.5Z_\odot$). This enables us to study the metallicity effect on the
magnetically split line blanketing that is due to the variation in the contribution of
individual elements with metallicity \citep[e.g.,][]{vikolamet}. The mass-loss
rates given in Table~\ref{ohvezpar} do not account for the dynamical effects of
the magnetic field. Therefore, these values in fact correspond to $\dot M_{B=0}$
rates that need to be further corrected to obtain the local mass flux that accounts
for dynamical effects \citep{owoan}.

It follows from the predicted mass-loss rates in Table~\ref{ohvezpar} that Zeeman
splitting has a negligible effect on the radiative force and on the wind
mass-loss rates. The relative change in mass-loss rates is on the order of
a few percent for magnetic field strengths of up to 10~kG. The mass-loss rate
decreases by about 10\%\  for the strongest magnetic field considered,
100~kG, which surpasses any magnetic field ever detected on the surfaces
of OB stars, however \citep[e.g.,][]{wadngc,grunmimes}. The decrease can be explained as
a result of line broadening, which, as shown in the case of turbulent
broadening, leads to a decrease in mass-loss rate \citep{cmf1}. For the
strongest magnetic fields we considered, the implicit assumption that the magnetic
splitting of the energy levels is small compared to the fine-structure splitting
may not be appropriate. For such fields a more general approach describing the so-called Paschen–Back effect should be used \citep[e.g.,][]{khalan}. This does not significantly affect the general results, however.

The magnetically split line blanketing is important only if the line shifts are
comparable with the line broadening. In our models we only assume thermal
broadening\footnote{We can neglect other types of broadening for
our purpose because, for example, in stars with strong magnetic fields macroturbulent broadening can be neglected because subsurface
convection is likely inhibited in strong magnetic fields \citep{sundin}.}, in which case
Eq.~\eqref{zeemstep} gives the condition for the minimum magnetic field strength,
\begin{equation}
\label{becko}
B=\frac{4\pi m_\text{e}c}{e\lambda_0g_l}\sqrt{\frac{2kT}{m}}=77\,\text{kG}
\zav{\frac{\lambda_0}{1000\,\text{\AA}}}^{-1}
\zav{\frac{T}{10^4\,\text{K}}}^{1/2}
\zav{\frac{m}{m_\text{H}}}^{-1/2},\end{equation}
assuming $g_l=g_u=1.2$ and $\Delta M=1$. Here $m$ is the atomic mass and
$m_\text{H}$ is the hydrogen atom mass. Eq.~\eqref{becko} shows that a magnetic field
with a strength of about 100\,kG is needed to affect the line force. Such a
magnetic field is higher than the upper limit of magnetic fields that  have been observed in
nondegenerate stars. This also explains why we did not find any strong effect
of the magnetic field on the line force.

The magnetic line splitting exceeds the Doppler shift that is connected with the
radial wind motion for magnetic fields that are stronger than about 1\,MG. Such strong
fields are typically found in some white dwarfs \citep[see][for a
review]{ab}.  In this case, the magnetically split lines behave independently and
do not interact with each other. Consequently, a stronger radiative
force and higher mass-loss rates can be expected. This might have implications for
hot ($T_\text{eff}\gtrsim100\,\text{kK}$) magnetic white dwarfs that have
winds (Krti\v cka et al., in preparation).

We did not find any strong flux variability that would be due to the magnetically split line
blanketing. The typical flux changes in the optical region at $5500\,$\AA\
correspond to magnitude variations of about $10^{-4}$\,mag.
Consequently, we do not expect any strong rotationally modulated flux
variability in magnetic O stars that would be purely due to the Zeeman splitting. A similar
result was obtained in magnetic main-sequence BA stars, where the magnetic field
only affects emergent fluxes in strongly overabundant atmospheres
\citep{zeeman-paper2}. The observed light variability in magnetic O stars
\citep{koeyer,nazmagmm} is therefore due to other processes, such as wind
blanketing that is modulated by the tilt of the magnetic field and stellar rotation
\citep{hd191612} or due to light absorption in a magnetically confined
circumstellar environment \citep{wahot,melimag}.

\section{Conclusions}

We studied the effect of line splitting that is due to the magnetic field (Zeeman
effect) on the wind properties in massive stars. We used our own numerical wind
code with CMF radiative transfer and NLTE level populations to estimate the
influence of the Zeeman splitting on the line force. We showed that for the
magnetic fields that are typically found in OB stars, the Zeeman splitting has a
negligible influence on the line force and also on the wind mass-loss rates and
terminal velocities. The line splitting only affects the radiative force for
magnetic fields that are stronger than about 100\,kG. We found only very weak flux
variability that is due to the magnetically split line blanketing. We conclude that only
dynamical effects connected with a magnetic field have a strong effect on the
mass-loss rate. These effects were deliberately neglected here because they
were studied using MHD models in detail, and we aimed at understanding of the
effect of the line splitting.

\section*{Acknowledgements}
The author thanks O.~Kochukhov for discussing the problem and the
anonymous referee for constructive comments. This work was supported by grant GA
\v{C}R 16-01116S. Access to computing and storage facilities owned by parties
and projects contributing to the National Grid Infrastructure MetaCentrum
provided under the program "Projects of Large Research, Development, and
Innovations Infrastructures" (CESNET LM2015042) is greatly appreciated.

\end{document}